\documentclass[onecolumn,noshowpacs,nofootinbib,amsmath,amssymb,preprint]{revtex4}

\usepackage{dcolumn}
\usepackage{graphicx}
\usepackage{bm}

\begin{document}

\title{MANIFESTATION OF THE $\bm{P}$-WAVE DIPROTON RESONANCE IN SINGLE-PION PRODUCTION \\ IN $\bm{pp}$ COLLISIONS}

\author{\firstname{M.~N.}~\surname{Platonova}}
\email{platonova@nucl-th.sinp.msu.ru}
\author{\firstname{V.~I.}~\surname{Kukulin}}
\email{kukulin@nucl-th.sinp.msu.ru}
\affiliation{Skobeltsyn Institute of Nuclear Physics, \\
Lomonosov Moscow State University, Moscow 119991, Russia}
%
%
\begin{abstract}
It is demonstrated that many important features of single-pion
production in $pp$ collisions at intermediate energies ($T_p
\simeq 400$--$800$ MeV) can naturally be explained by supposing
excitation of intermediate diproton resonances in $pp$ channels
${}^1D_2$, ${}^3F_3$ and ${}^3P_2$, in addition to conventional
mechanisms involving an intermediate $\Delta$-isobar. We predict
for the first time the crucial role of the ${}^3P_2$ diproton
resonance, found in recent experiments on the single-pion
production reaction $pp \to pp({}^1S_0) \pi^0$, in reproducing the
proper behavior of spin-correlation parameters in the reaction $pp
\to d \pi^+$ which were poorly described by conventional
meson-exchange models to date. The possible quark structure of the
$P$-wave diproton resonances is also discussed.
\end{abstract}


\maketitle

\newpage

\section{Introduction. Brief historical excursus}
\label{sec1}

The activity in searching for dibaryon resonances in 1980s was
motivated by the success of MIT-bag models in prediction of
dibaryon states~\cite{Jaffe77,Aerts78,Matveev78}. Numerous
experiments on $\vec{p}\vec{p}$ elastic scattering done at the
same time revealed possible existence of a series of diproton
resonances with masses in the range $2.1$--$2.9$ GeV
and total widths $100$--$200$
MeV~\cite{Auer77,Auer78,Biegert78,Auer82,Auer89}. Further studies
established that these resonances are mainly of inelastic nature
and seen primarily in inelastic channels like $pp \to d \pi^+$,
$pp \to pn \pi^+$, etc.~\cite{Locher86}. Using different data
sets, a few groups performed partial-wave analyses (PWA) of $pp$
and $\pi^+ d$ elastic scattering and the $pp \leftrightarrow d
\pi^+$
reaction~\cite{Bhandari81,Arndt92,Kravtsov83,Strakovsky84,Hoshizaki78,Hoshizaki79}
and found resonance poles in the ${}^1D_2$, ${}^3F_3$, ${}^3P_2$,
${}^1G_4$ and other $NN$ channels. However some authors suggested
the observed singularities to be related to the so-called
pseudoresonances (see, e.g.,~\cite{Simonov79}), which means rather
generation of a resonance in a subsystem instead of the true
diproton resonance in a whole interacting system. In case of $pp$
scattering at energies $T_p \simeq 600$ MeV, the pseudoresonance
implies an intermediate $\Delta$-isobar generation coupled
strongly to the rest nucleon. Thus, the resonance behavior of the
$pp$ elastic and inelastic scattering amplitudes is basically
associated with the nearby $N\Delta$ thresholds in the respective
partial waves~\cite{Niskanen82,Strakovsky91}. Similar discussions
about dibaryon resonances near the $\Delta\Delta$ threshold were
very active for the last three decades, at least (see,
e.g.,~\cite{Valcarce01}).

This rather indefinite situation began to change only in recent
years when an experimental group using the $4\pi$ detector WASA
installed at the COSY facility (Juelich) together with the SAID
Data Analysis Center announced~\cite{Adl11,Adl14-el} the discovery
of an $I(J^P)=0(3^+)$ dibaryon resonance $d^*$ with a mass
$M_{d^*} \simeq 2.38$ GeV in the ${}^3D_3$--${}^3G_3$ channels of
$NN$ system both in $2\pi$-production reactions and $\vec{n}p$
elastic scattering. This resonance called an ``inevitable
dibaryon''~\cite{Goldman89} had been searched for 50 years since
its first prediction by Dyson and Xuong still in
1964~\cite{Dyson64}. Remarkably, the $d^*$ dibaryon was predicted
in~\cite{Dyson64} to belong to the same SU(3) multiplet as the
deuteron --- the lowest isoscalar dibaryon, while the ${}^1D_2$
dibaryon was predicted to belong to the same SU(3) multiplet as
the singlet deuteron --- the lowest isovector dibaryon. Very
recently another experimental group which uses the forward
detector ANKE at the COSY facility have received an
evidence~\cite{Komarov16} of the ${}^3P_2$ and ${}^3P_0$ diproton
resonances with a mass $M_D \simeq 2.2$ GeV in the reaction $pp
\to (pp)_{0}\pi^0$, where $(pp)_{0}$ means the ${}^1S_0$ singlet
deuteron near-threshold state.

From the theoretical side, the calculations~\cite{Gal13,Gal14}
within the framework of rigorous three-body $\pi NN$ and $\pi
N\Delta$ models revealed a robust ${}^1D_2$ dibaryon resonance
pole near the $N \Delta$ threshold and also a ${}^3D_3$ resonance
pole near (below) the $\Delta \Delta$ threshold. Furthermore, the
recent quark model studies of the $d^*$ dibaryon strongly support
its unconventional nature as being a genuine six-quark state
rather than just a $\Delta$--$\Delta$ bound state. Indeed, the
observed width and decay properties of this resonance can be
explained only if one assumes that it is dominated by a
``hidden-color'' six-quark
configuration~\cite{Bash13,Huang14,Dong15}. The ``hidden-color''
six-quark states are a rigorous first-principle prediction of
SU(3) color gauge theory~\cite{Brodsky83,Brodsky86}. We also cite
in this connection the recent issue of CERN
Courier~\cite{CernCourier14} in section ``News. New particles'':
``COSY confirms existence of six-quark states''. So, in light of
these new findings, one may hope that the long-term dispute
between the supporters of the near-threshold singularities in
$N\Delta$ and $\Delta\Delta$ channels and the apologists of the
true dibaryon resonances will shortly come to its completion.

If to consider the dispute between two above alternatives from the
general physical point of view, we should say that as was
recognized still long ago by Baz'~\cite{Baz59} who developed
Wigner's ideas~\cite{Wigner48} on the near-threshold cross section
singularities in the field of nuclear reactions, there should be
(in majority of nuclei) a strong correlation between the position
of a threshold for some channel $B+C$ in a nucleus $A=B+C$ and the
near-threshold energy levels with appropriate quantum numbers. It
is because the fragments $B$ and $C$ can move far apart near the
channel threshold keeping thereby their identity, so that, a
near-threshold bound (or resonance) state should emerge very
likely. A careful inspection~\cite{Serov63} of the well-known
nuclear level tables~\cite{Ajzenberg59} actually confirmed the
close correlation between the channel thresholds and the nearby
bound states in many nuclei (e.g., ${}^{12}{\rm C}^* \to {}^8{\rm
Be} + \alpha$, ${}^{16}{\rm O}^* \to {}^{12}{\rm C}^* + \alpha$,
etc.). Hence, it can be supposed quite naturally that there is a
strong correlation between thresholds and the nearby bound (or
resonance) states also in hadronic
physics~\cite{Hanhart10,Hyodo13}. A good example may be the Roper
resonance $N^*(1440)$, its average pole mass\footnote{One should
bear in mind that a double-pole structure with two almost
degenerate poles was found for the Roper resonance
in~\cite{Arndt85} and recently confirmed by ANL-Osaka and Juelich
groups.} being $M_{\rm pole} \simeq 1365$ MeV~\cite{PDG14}. In
fact, it has been found
experimentally~\cite{Sarantsev08,Skorodko08} that the very large
(or even dominating) decay mode for the Roper resonance is the
light scalar $\sigma$-meson emission (with $m_{\sigma} \simeq
400$--$500$ MeV), so that, the Roper can be treated as a
near-threshold state in the $\sigma + N$ channel~\cite{Obukh11}.
The recent Faddeev calculations for ``meson-assisted
dibaryons''~\cite{Gal16} seem to confirm the general correlation
between thresholds and the nearby bound (or resonance) states in
the dibaryon field as well.


Moreover, QCD does not forbid existence of multiquark states near
thresholds or elsewhere. Recent experimental discoveries of the
tetra- and pentaquarks~\cite{Aaij14,Aaij15} have confirmed
existence of exotic multiquark states in general. So, in view of
all these new achievements, studying the properties of multiquark
states and their manifestation in the basic hadronic processes has
become of particular importance now.

The present paper is dedicated to study of manifestation of
diproton (dibaryon) resonances in single-pion production in $pp$
collisions in the GeV region. The main emphasis will be given to
the basic pion-production reaction $pp \to d \pi^+$ at energies
$T_p \simeq 400$--$800$ MeV where a few PWA as well as a rich set
of experimental data exist. In the work~\cite{NPA16} we elaborated
a model which combines two dominating conventional mechanisms of
this reaction, i.e., one-nucleon exchange and an intermediate
$\Delta$ excitation, with the resonance mechanisms based on
intermediate dibaryons excitation. By re-examining the
conventional $\Delta$-excitation mechanism, we have shown its
strong sensibility to the short-range cut-off parameters $\Lambda$
in meson-baryon vertices, especially in the $\pi N\Delta$ vertex.
Thus, when using ``soft'' cut-off parameters which naturally arise
from description of $\pi N$ elastic scattering in the $\Delta$
region, the conventional meson-exchange mechanisms give a strong
underestimation for the partial and total $pp \to d \pi^+$ cross
sections. So, we have shown that the significant contribution
should come from other sources (of the short-range nature), and
that excitation of intermediate dibaryons in the dominant partial
waves ${}^1D_2P$ and ${}^3F_3D$ of the reaction $pp \to d \pi^+$
can really give this lacking contribution.

To our knowledge, the only attempt (besides the PWA) to describe
the reaction $pp \to d \pi^+$ at energies $T_p \simeq 400$--$800$
MeV including dibaryon resonances was made previously in the
works~\cite{Kamo79,Kamo80}. The authors used essentially the same
model as we did in~\cite{NPA16}, but with a more sophisticated
treatment of $NN \to N\Delta$ amplitudes, and came to a conclusion
similar to ours, that the conventional mechanisms give only a half
of the total cross section. Then, by fitting the parameters of six
hypothetical dibaryon resonances to the existing experimental
data, they also revealed importance of two dibaryon resonances,
${}^1D_2$ and ${}^3F_3$, in reproducing the total and differential
cross sections and also the proton analyzing power. However, the
masses of resonances other than ${}^1D_2$ and ${}^3F_3$ were found
in their analysis to be too low, and their widths too narrow.
Besides that, their model calculations could not reproduce the
spin-correlation parameters properly. On the contrary, we fit the
results of the most recent PWA~\cite{Arndt93,Oh97} rather than
experimental data, and include dibaryons only in three dominant
partial waves, ${}^1D_2P$, ${}^3F_3D$ and (in the present paper)
${}^3P_2D$, where the resonance behavior of the amplitudes is well
established. Thus, we can extract the dibaryon parameters more
precisely and judge on the role of individual resonances in $pp
\to d\pi^+$ observables.

In the present study, we focus on differential observables of the
reaction $pp \to d\pi^+$ and on the role of the $P$-wave diproton
resonance which can be excited in the ${}^3P_2D$ partial wave of
the reaction. The ${}^3P_2$ diproton resonance found previously in
the PWA of $pp$ elastic scattering~\cite{Arndt87,Higuchi91} and
confirmed in a recent experiment on the reaction $pp \to
(pp)_{0}\pi^0$~\cite{Komarov16} received much less attention in
literature than the ${}^1D_2$ and ${}^3F_3$ resonances. As we will
show in the paper, the ${}^3P_2$ dibaryon, though giving a small
contribution ($< 10$\%) to the total $pp \to d\pi^+$ cross
section, is very important for reproducing the differential
observables, especially spin-correlation parameters, which have
been poorly described by the conventional meson-exchange models up
to date~\cite{Lamot87,Niskanen84,Grein84} and also by a
model~\cite{Kamo79,Kamo80} which included dibaryon resonances.

The structure of the paper is following. In Sect.~\ref{sec2} the
working model for treatment of the reaction $pp \to d\pi^+$ with
both intermediate $\Delta$'s and dibaryons is formulated. In
Sect.~\ref{sec3} the description of the partial cross sections in
the dominant partial waves ${}^1D_2P$, ${}^3F_3D$ and ${}^3P_2D$
and also of the total cross section in a broad energy range is
given. Sect.~\ref{sec4} is devoted to discussion of differential
cross section, as well as vector analyzing powers and
spin-correlation parameters, at energy $T_p = 582$ MeV. In
Sect.~\ref{sec5} the basic results attained in the paper are
summarized and discussed.

\section{Theoretical model}
\label{sec2}

\begin{figure}[!ht]
\begin{center}
\resizebox{0.7\textwidth}{!}{\includegraphics{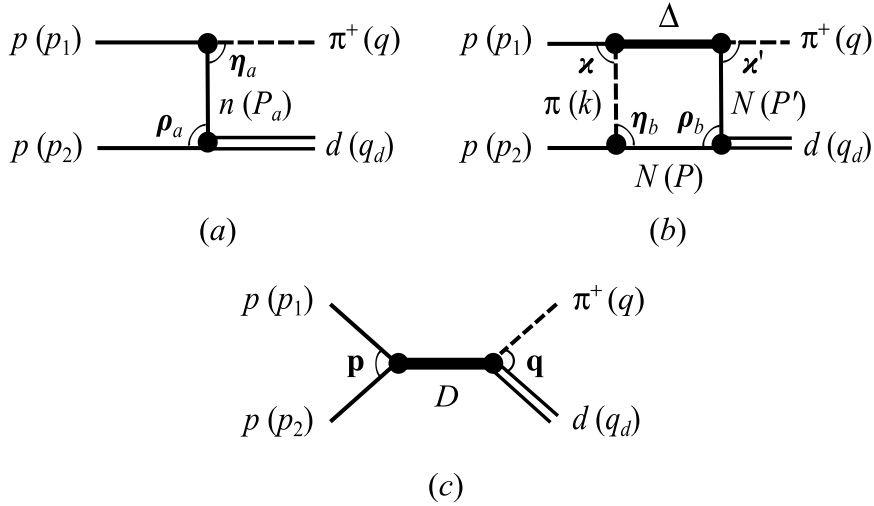}}
\end{center}
\caption{Diagrams illustrating three basic mechanisms for the
reaction $pp \to d \pi^+$: one-nucleon exchange ($a$),
intermediate $\Delta$-isobar excitation ($b$), and intermediate
dibaryon resonance excitation ($c$). The 4-momenta of the
particles are shown in parentheses, and 3-momenta in pair
center-of-mass systems are denoted by bold face.} \label{fig1}
\end{figure}

In this section, we briefly outline our model formalism for the
reaction $pp \to d \pi^+$. The details can be found in
Ref.~\cite{NPA16}. The model includes three basic mechanisms
depicted in Fig.~\ref{fig1}. Two conventional mechanisms, i.e.,
one-nucleon exchange and excitation of the intermediate $N\Delta$
system by the $t$-channel pion exchange are shown in
Figs.~\ref{fig1}~$(a)$ and $(b)$, respectively. Further on, we
will refer to these mechanisms as ONE and $N\Delta$. An excitation
of the intermediate $\Delta$ isobar through the $\rho$-meson
exchange was also often considered in the
literature~\cite{Brack77}, but such a mechanism contributes
significantly only when choosing very high cut-off parameters in
the meson-baryon form factors. Here, we choose the low values for
the cut-off parameters $\Lambda < 1$ GeV (reasons for this will be
given below), for which the contribution of the $\rho$-exchange
mechanism is very small.

In a standard approximation of the spectator
nucleon~\cite{Brack77}, the helicity amplitudes corresponding to
the mechanisms ONE and $N\Delta$ take the form:
\begin{equation}
\mathcal{M}^{(\rm{ONE})}_{\lambda_1,\lambda_2;\lambda_d} =
-\sqrt{2} (2m)^{3/2} \chi^{\dag}(\lambda_2)i \sigma_2 \times
\Psi^*_{d}(\bm{\rho_a},\lambda_d)F_{\pi
NN}(\eta_a)(\bm{\sigma\eta_a}) \chi(\lambda_1), \label{onem}
\end{equation}
\begin{equation*}
\mathcal{M}^{(N\Delta)}_{\lambda_1,\lambda_2;\lambda_d} =
-4\sqrt{2}/3 (2m)^{1/2} \chi^{\dag}(\lambda_2)i \sigma_2 \, \int
\frac{d^3 P}{(2\pi)^3} \frac{F_{\pi
NN}(\eta_b)(\bm{\sigma\eta_b})}{w_{\pi}^2-m_{\pi}^2+i0}
\end{equation*}
\begin{equation}
\times \Psi^*_{d}(\bm{\rho_b},\lambda_d)
\sqrt{\frac{\Gamma_{\Delta}(\varkappa)\Gamma_{\Delta}(\varkappa')}{\varkappa^3
\varkappa'^3}} \frac{16 \pi W_{\Delta}^2 (\bm{\varkappa\varkappa'}
 + i \frac{\bm{\sigma}}{2}\bm{\varkappa\times\varkappa'})}{W_{\Delta}^2 - M_{\Delta}^2 + i
W_{\Delta}\Gamma_{\Delta}(W_{\Delta})}\chi(\lambda_1),
\label{ndem}
\end{equation}
where $w_{\pi}^2=k^2$, and the $\Delta$-isobar width is related to
the vertex function $F_{\pi N\Delta}$ as
\begin{equation}
\Gamma_{\Delta}(\varkappa) = \frac{\varkappa^3 m}{6\pi
W_{\Delta}}F_{\pi N\Delta}^2(\varkappa). \label{gf}
\end{equation}

To calculate the spin structure of the amplitudes, it is
convenient to write the deuteron wave function as
\begin{equation}
\label{dwf} \Psi_{d}(\bm{\rho},\lambda_d) = \bm{\sigma}{\bf
E}(\bm{\rho},\lambda_d),
\end{equation}
where we introduced the vector
\begin{equation}
\label{e-vect} {\bf E}(\bm{\rho},\lambda_d) =
u(\rho)\bm{\varepsilon}(\lambda_d)  +  \frac{w(\rho)}{\sqrt{2}}
\!\! \left(\bm{\varepsilon}(\lambda_d) -
\frac{3\bm{\rho}(\bm{\rho\varepsilon}(\lambda_d))}{\rho^2}\right).
\end{equation}
Here, $\bm{\varepsilon}(\lambda_d)$ is the standard deuteron
polarization vector, $u$ and $w$ are the $S$- and $D$-wave
components of the d.w.f. normalized as $\int d^3 \rho \left(u^2 +
w^2\right)/ {(2\pi)^3} = 1$.

The helicity amplitudes antisymmetrized over two initial protons
take the form
\begin{equation}
\mathcal{M}^{(s)}_{\lambda_1,\lambda_2;\lambda_d}(\theta) =
\mathcal{M}_{\lambda_1,\lambda_2;\lambda_d}(\theta) \,+\,
(-1)^{\lambda_d}\mathcal{M}_{\lambda_2,\lambda_1;\lambda_d}(\pi-\theta).
\end{equation}
Overall, there are 6 independent helicity amplitudes in the
reaction $pp \to d \pi^+$~\cite{Grein84}:
\begin{equation*}
\Phi_1 = \mathcal{M}^{(s)}_{\frac{1}{2},\frac{1}{2};1}, \quad
\Phi_2 = \mathcal{M}^{(s)}_{\frac{1}{2},\frac{1}{2};0}, \quad
\Phi_3 = \mathcal{M}^{(s)}_{\frac{1}{2},\frac{1}{2};-1},
\end{equation*}
\begin{equation}
\Phi_4 = \mathcal{M}^{(s)}_{\frac{1}{2},-\frac{1}{2};1}, \quad
\Phi_5 = \mathcal{M}^{(s)}_{\frac{1}{2},-\frac{1}{2};0}, \quad
\Phi_6 = \mathcal{M}^{(s)}_{\frac{1}{2},-\frac{1}{2};-1}.
\end{equation}

For comparison of the theoretical results with the PWA data and
for studying the contributions of the intermediate dibaryon
resonances, it is convenient to deal with the partial-wave
amplitudes, which are expressed through the helicity ones via the
standard formulas given by Jacob and Wick~\cite{Jacob59}. The
dominant partial-wave amplitudes in a broad energy range including
the region of $\Delta$ excitation, as was shown by $\pi^+d \to pp$
PWA (see, e.g., Fig.~5$b$ in Ref.~\cite{Arndt93}), are ${}^1D_2P$,
${}^3F_3D$ and ${}^3P_2D$ (with decreasing magnitude). The
explicit formulas for these amplitudes are
\begin{equation}
A({}^1D_2P) = \frac{1}{2}\sqrt{\frac{3}{5}}\left(\Phi_1^{(2)} +
\Phi_3^{(2)}\right) + \frac{1}{\sqrt{5}}\Phi_2^{(2)}, \label{ad}
\end{equation}
\begin{equation}
A({}^3F_3D) = -\frac{2}{\sqrt{7}}\Phi_4^{(3)} -
\frac{1}{2}\sqrt{\frac{6}{7}} \Phi_5^{(3)}, \label{af}
\end{equation}
\begin{equation}
A({}^3P_2D) = \sqrt{\frac{1}{10}}\left(\Phi_1^{(2)} -
\Phi_3^{(2)}\right) + \sqrt{\frac{3}{5}} \Phi_4^{(2)}, \label{ap}
\end{equation}
where
\begin{equation}
\Phi_i^{(J)} =
\int\limits_{-1}^{1}d^{(J)}_{\lambda_1-\lambda_2,-\lambda_d}(x)\Phi_i(x)dx,
\quad x = \rm{cos}(\theta). \label{phij}
\end{equation}

For the amplitude corresponding to excitation of an intermediate
dibaryon resonance (see Fig.~\ref{fig1}$(c)$), it is convenient to
start from the partial-wave representation. The respective
amplitude is
\begin{equation}
A^{(D)}({}^{2S+1}L_JL_{\pi}) = -\frac{8\pi
s}{\sqrt{pq}}\frac{\sqrt{2
\Gamma_{i}(s)\,\Gamma_{f}(s)}}{s-M_{D}^2+i\sqrt{s}\Gamma_{D}(s)},
\label{adib}
\end{equation}
where $p = \left(s - 4m^2\right)^{1/2}/2$ and $q = \left[\left(s -
m_{\pi}^2 - m_d^2\right)^2 - 4 m_{\pi}^2
m_d^2\right]^{1/2}/2\sqrt{s}$ are the moduli of the proton and the
pion c.m.s. momenta, respectively. The factor 2 before the
incoming width $\Gamma_{i}(s)$ was introduced to account for two
identical protons in the initial state.

For the incoming width $\Gamma_i(s) \equiv \Gamma_{D \to pp}(s)$,
we used the Gaussian parameterization which follows from the $D
\to NN$ form factor parameterization employed in the dibaryon
model for $NN$ interaction~\cite{JPG01K,IJMP02K}:
\begin{equation}
\Gamma_{i}(s) = \Gamma_{i} \left(\frac{p}{p_0}\right)^{2L+1} {\rm
exp}\left(-\frac{p^2-p_0^2}{\alpha_{pp}^2}\right), \label{gdnn}
\end{equation}
where $p_0$ is the value of the $pp$ relative momentum at
$\sqrt{s} = M_D$.

For the outgoing width $\Gamma_f(s) \equiv \Gamma_{D \to \pi^+
d}(s)$, we employed the parameterization analogous to that for the
$\Delta \to \pi N$ width (cf.~(\ref{gf}) with a monopole form
factor~(\ref{fpind})):
\begin{equation}
\Gamma_{f}(s) = \Gamma_{f} \left(\frac{q}{q_0}\right)^{2L_{\pi}+1}
\left(\frac{p_0^2-\Lambda_{\pi d}^2}{p^2-\Lambda_{\pi
d}^2}\right)^{L_{\pi}+1}, \label{gdpid}
\end{equation}
where $q_0$ is the value of the $\pi d$ relative momentum at
$\sqrt{s} = M_D$. This parameterization was proposed for $\pi N$
and $KN$ elastic scattering still in~\cite{Jackson64,Anisovich69}
and then applied for the $\pi^+ d \to pp$ PWA
in~\cite{Strakovsky84}. The same energy dependence was assumed
here also for the total dibaryon width $\Gamma_{D}(s)$, since, due
to the high inelasticity of dibaryon resonances, the incoming
width $\Gamma_{D \to pp}$ is only a small fraction (ca. $10$\%) of
the total width~\cite{Strakovsky91}.

By using Eq.~(\ref{adib}) and the Jacob--Wick
formulas~\cite{Jacob59} which are an inversion of
Eqs.~(\ref{ad})--(\ref{ap}), one can find the contributions from
intermediate dibaryons to the helicity amplitudes $\Phi_i$ ($i=1,
\ldots 6$). One should also note that $\Phi_6^{(J)}=\Phi_4^{(J)}$
for odd $J$ and $\Phi_6^{(J)}=-\Phi_4^{(J)}$ for even
$J$~\cite{Grein84}. Then the respective helicity amplitudes, when
three dibaryon resonances are taken into account, take the form
\begin{equation*}
\Phi^{(D)}_1 = \left(\frac{\sqrt{15}}{2} A^{(D)}({}^1D_2P)  +
\sqrt{\frac{5}{2}} A^{(D)}({}^3P_2D)\right) \, d^{(2)}_{0,-1}(x),
\end{equation*}
\begin{equation*}
\Phi^{(D)}_2 = \sqrt{5} A^{(D)}({}^1D_2P) \, d^{(2)}_{0,0}(x),
\end{equation*}
\begin{equation*}
\Phi^{(D)}_3 = \left(\frac{\sqrt{15}}{2} A^{(D)}({}^1D_2P)  -
\sqrt{\frac{5}{2}} A^{(D)}({}^3P_2D)\right) \, d^{(2)}_{0,1}(x),
\end{equation*}
\begin{equation*}
\Phi^{(D)}_4 = -\sqrt{7} A^{(D)}({}^3F_3D)d^{(3)}_{1,-1}(x) +
\frac{\sqrt{15}}{2} A^{(D)}({}^3P_2D)d^{(2)}_{1,-1}(x),
\end{equation*}
\begin{equation*}
\Phi^{(D)}_5 = -\sqrt{\frac{21}{2}} A^{(D)}({}^3F_3D) \,
d^{(3)}_{1,0}(x),
\end{equation*}
\begin{equation}
\Phi^{(D)}_6 = -\sqrt{7} A^{(D)}({}^3F_3D)d^{(3)}_{1,1}(x) -
\frac{\sqrt{15}}{2} A^{(D)}({}^3P_2D)d^{(2)}_{1,1}(x).
\end{equation}
One can see that the ${}^1D_2P$ amplitude gives the dominant
contribution to the helicity amplitudes $\Phi_1$--$\Phi_3$, while
the ${}^3F_3D$ amplitude gives the dominant contribution to
$\Phi_4$--$\Phi_6$. At the same time, the ${}^3P_2D$ amplitude
introduces corrections to both sets of helicity amplitudes. As
will be shown in Sect.~\ref{sec4}, these corrections, though being
rather small in magnitude, turn out to be crucial for polarization
observables in the $pp \to d \pi^+$ reaction.

The partial cross sections are expressed through the partial-wave
amplitudes as follows:
\begin{equation}
\sigma({}^{2S+1}L_JL_{\pi}) = \frac{(2J+1)}{64\pi
s}\frac{q}{p}\left|A({}^{2S+1}L_JL_{\pi})\right|^2.
\end{equation}

Further, we give the expressions for observables in terms of six
helicity amplitudes $\Phi_i$ ($i=1, \ldots 6$), using the
notations of Ref.~\cite{Grein84} (apart from a $2m$ factor in the
amplitudes normalization), with the signs of polarization
observables given in Madison convention. A different notation for
amplitudes and observables can be found in, e.g.,~\cite{Arndt93}.

For the total cross section, one has
\begin{equation}
\sigma(pp \to d \pi^+) = \frac{1}{64\pi
s}\frac{q}{p}\int\limits_{-1}^{1}\sum\limits_{i=1}^{6}\left|\Phi_i(x)\right|^2
dx.
\end{equation}

The following expressions hold for the differential cross section:
\begin{equation}
\frac{d\sigma}{d\Omega}(pp \to d \pi^+) = \frac{1}{64\pi^2
s}\frac{q}{p}\frac{1}{4}\Sigma, \quad \Sigma = 2
\sum\limits_{i=1}^{6}\left|\Phi_i\right|^2,
\end{equation}
for proton and deuteron vector analyzing powers:
\begin{equation}
A_{y0} = 4 {\rm Im} \left(\Phi_1^*\Phi_6 + \Phi_3^*\Phi_4 -
\Phi_2^*\Phi_5\right)\Sigma^{-1},
\end{equation}
\begin{equation}
iT_{11} = -\sqrt{6} {\rm Im}
\left[\left(\Phi_1^*-\Phi_3^*\right)\Phi_2 +
\left(\Phi_4^*-\Phi_6^*\right)\Phi_5\right]\Sigma^{-1},
\end{equation}
and for proton-proton spin-correlation parameters:
\begin{equation}
A_{xx} = \left[4 {\rm Re} \left(\Phi_1^*\Phi_3 -
\Phi_4^*\Phi_6\right) + 2\left|\Phi_5\right|^2 -
2\left|\Phi_2\right|^2 \right]\Sigma^{-1},
\end{equation}
\begin{equation}
A_{yy} = \left[4 {\rm Re} \left(\Phi_1^*\Phi_3 +
\Phi_4^*\Phi_6\right) - 2\left|\Phi_5\right|^2 -
2\left|\Phi_2\right|^2 \right]\Sigma^{-1},
\end{equation}
\begin{equation}
A_{zz} = -2\left(\left|\Phi_1\right|^2 + \left|\Phi_2\right|^2 +
\left|\Phi_3\right|^2 - \left|\Phi_4\right|^2 -
\left|\Phi_5\right|^2 - \left|\Phi_6\right|^2\right)\Sigma^{-1},
\end{equation}
\begin{equation}
A_{xz} = 4 {\rm Re} \left(\Phi_1^*\Phi_6 +
\Phi_3^*\Phi_4-\Phi_2^*\Phi_5 \right)\Sigma^{-1}.
\end{equation}
There are also deuteron tensor analyzing powers and
spin-correlation parameters for proton and deuteron and two
protons and deuteron~\cite{Arndt93,Grein84}. However, experimental
data exist for the above-defined observables only, so, in the
present paper, we restrict our calculations to these observables.

The meson-baryon vertex functions $F_{\pi NN}$ and $F_{\pi
N\Delta}$ were parameterized in a monopole form\footnote{The
cut-off parameters $\tilde{\Lambda}$ and $\tilde{\Lambda}_*$ in
Eq.~(\ref{fpind}) were marked by a tilde sign to distinguish them
from parameters used in a more familiar monopole vertex
parameterization which follows from Eq.~(\ref{fpind}) when only
pion is off-shell (see Ref.~\cite{NPA16} for details).}
\begin{equation}
\label{fpind} F_{\pi NN}(p,\tilde{\Lambda}) =
\frac{f}{m_{\pi}}\frac{p_0^2 + \tilde{\Lambda}^2}{p^2 +
\tilde{\Lambda}^2}, \quad F_{\pi N \Delta}(p,\tilde{\Lambda}_*) =
\frac{f_*}{m_{\pi}}\frac{p_{0}^2 + \tilde{\Lambda}_*^2}{p^2 +
\tilde{\Lambda}_*^2},
\end{equation}
where $p^2$ is a modulo squared of the $\pi$--$N$ relative
momentum (i.e., the pion momentum in the $\pi N$ c.m.s.) and
$p_0^2$ corresponds to the situation when all three particles are
real, i.e., located on their mass shells (so, $p_0^2$ is positive
for $\pi N\Delta$ and negative for $\pi NN$ vertex). The coupling
constants in Eq.~(\ref{fpind}) have been taken to be $f = 0.97$
and $f_* = 2.17$. In this case, one has $f^2/4\pi=0.075$, and the
above value for $f_*$ was derived from the total width of the
$\Delta$ isobar $\Gamma_{\Delta} = 117$ MeV as given by the
Particle Data Group~\cite{PDG14}.

As was argued in~\cite{NPA16}, the main advantage of the above
vertex parametrization is that it admits a straightforward
off-shell continuation and describes the real and virtual
particles \textit{in a unified manner}. It does not require
introducing any additional parameters to account for the particles
leaving their mass shells. Hence, it can be used for consistent
description of processes involving on- and off-shell pions, i.e.,
$\pi N \to \pi N$, $NN \to \pi d$, elastic $NN$ scattering, etc.,
with the same cut-off parameters in meson-baryon vertices.
Moreover, the cut-off parameters $\Lambda$ in such a case do not
need to be fitted \textit{ad hoc} and can in general be found
directly from experimental data.

\begin{figure}[!ht]
\begin{center}
\resizebox{0.6\columnwidth}{!}{\includegraphics{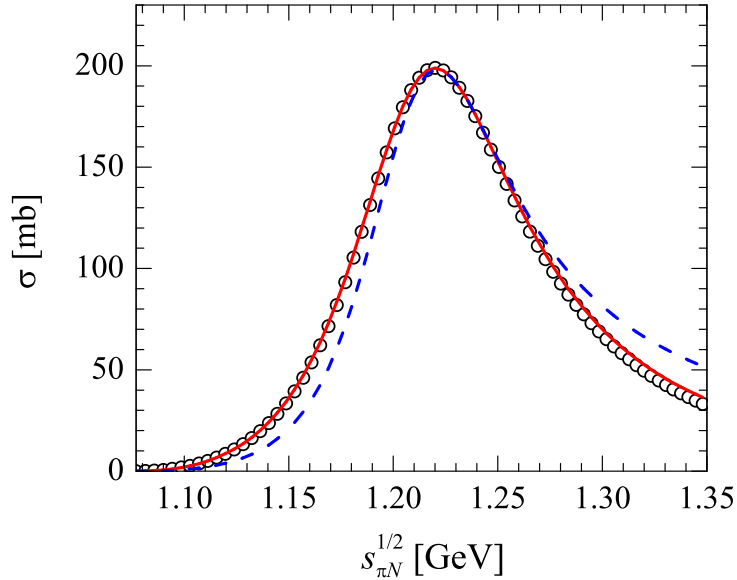}}
\end{center}
\caption{The cross section of $\pi N$ elastic scattering in the
$P_{33}$ partial wave. Solid and dashed lines show the
calculations in the isobar model with the $\pi N \Delta$ vertex in
the form (\ref{fpind}) and cut-off parameters $\tilde{\Lambda}_* =
0.3$ and $0.52$~GeV, respectively. Open circles correspond to the
PWA data (SAID, solution WI08~\cite{SAID}).} \label{fig2}
\end{figure}

Thus, the parameter $\tilde{\Lambda}_*$ in the $\pi N \Delta$
vertex can be found from empirical data on $\pi N$ elastic
scattering. From fitting the PWA (SAID) data~\cite{SAID} for the
$\pi N$-scattering $P_{33}$ partial cross section in a broad
energy range within the isobar model, we found $\tilde{\Lambda}_*
= 0.3$ GeV (see Fig.~\ref{fig2}). For the $\pi NN$ vertex, we have
chosen the value $\tilde{\Lambda} = 0.7$ GeV, which was used in a
number of previous calculations of single-pion production
reactions~\cite{Grein84,Uzikov88}. This value of $\tilde{\Lambda}$
is also consistent with the predictions of the lattice-QCD
calculations~\cite{Liu99,Erkol09}.

The detailed discussion on the choice of short-range cut-off
parameters and their strong impact on cross sections of the $pp
\to d\pi^+$ reaction can be found in our previous
work~\cite{NPA16}. It should be stressed here that the cut-off
values chosen in our calculations are much lower than those
traditionally used in the realistic $NN$-potential models. For
example, in the Bonn model~\cite{Machl87}, the minimal values,
which still allow a good description of $NN$-scattering phase
shifts up to $T_N = 350$ MeV, are $\Lambda \simeq \Lambda_* \simeq
1.3$ GeV (in the CD-Bonn model~\cite{Machl01} they are even
higher). Such very high cut-off parameters apparently lead to
increased meson-exchange contributions at short inter-nucleon
distances. On the other hand, results of the numerous quark-model
calculations agree, in general, that the parameters in
meson-baryon vertices should be essentially soft, i.e., $\Lambda <
1$ GeV (see, e.g.,~\cite{Koepf96} and references therein). In this
case, one should seek for some alternative short-range mechanisms
(such as formation of intermediate dibaryons) to describe the
processes involving high momentum transfers within the two-nucleon
system.

\section{Results for the partial and total cross sections}
\label{sec3}

Here we calculated partial and total cross sections for the
reaction $pp \to d\pi^+$ in the energy range $\sqrt {s} =
2.03$--$2.27$ GeV ($T_p \simeq 320$--$860$ MeV) using the above
formalism. Three dibaryon resonances generated in $pp$ channels
${}^1D_2$, ${}^3F_3$ and ${}^3P_2$ were included in calculations.
At the present stage, we restricted ourselves to accurate
description of three dominant partial-wave amplitudes and to
qualitative estimation of dibaryon contributions in these
amplitudes. We did not consider the possible dibaryons in small
amplitudes (such as ${}^1G_4$, ${}^3P_1$, ${}^1S_0$, etc.), since
the level of evidence for these dibaryons is less to date than for
${}^1D_2$, ${}^3F_3$ and ${}^3P_2$ ones, and also description of
the small amplitudes would require a more precise treatment of the
background meson-exchange processes. Besides that, increasing the
number of dibaryons would increase the number of model parameters
and thus complicate making reliable conclusions.

For consistency of our model, in calculations of conventional
mechanisms ONE and $N\Delta$, we used the deuteron wave function
(d.w.f.) derived in the dibaryon model for $NN$
interaction~\cite{JPG01K,IJMP02K}. This d.w.f. has been truncated
in the present study at high inter-nucleon momenta $p
> 350$ MeV (with keeping the overall normalization) to prevent an
unphysical rise of the differential cross section at large angles.
So, the results obtained here with such regularized d.w.f. turned
out to be very close to those with the conventional CD-Bonn
d.w.f.~\cite{Machl01}.

Our model calculations were compared to the results of the most
recent PWA (SAID, solution C500~\cite{SAID,Oh97}), which is a
coupled-channel analysis using $\pi^+ d \rightarrow pp$, $pp \to
pp$ and $\pi^+ d \to \pi^+ d$ experimental data. The dibaryon
parameters obtained by fitting the partial cross sections in three
dominant partial waves ${}^1D_2P$, ${}^3F_3D$ and ${}^3P_2D$ to
the PWA results are summarized in Table I. The relative phases
$\varphi$ between the resonance (dibaryon) and ``background'' (ONE
+ $N\Delta$) amplitudes were fixed as shown in the last column of
the table (these values coincide with the best-fit results up to
several degrees).

The dibaryon masses and widths obtained in our fit are generally
consistent with the previous
estimates~\cite{Bhandari81,Hoshizaki93,Yokosawa90}. It is
particularly important that the parameters of the ${}^3P_2$
resonance found here are in a very good agreement with those found
in a recent experimental work~\cite{Komarov16} from a global fit
of experimental data on the differential cross section and proton
analyzing power in the reaction $pp \to (pp)_0 \pi^0$, i.e., $M_D
= 2207 \pm 12$ and $\Gamma_D = 170 \pm 32$ MeV.

\begin{figure}[!ht]
\begin{center}
\resizebox{0.72\columnwidth}{!}{\includegraphics{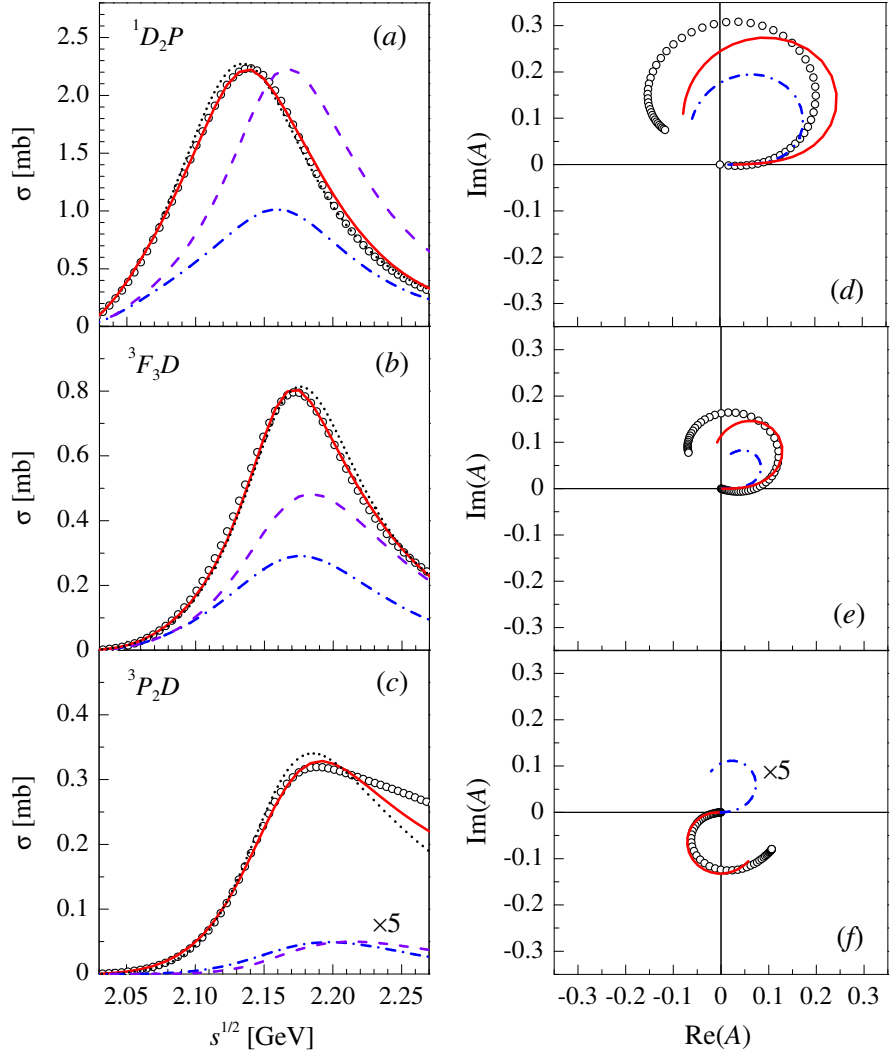}}
\end{center}
\caption{Left panel: partial cross sections of the reaction $pp
\to d \pi^+$ in the dominant partial waves ${}^1D_2P$ ($a$),
${}^3F_3D$ ($b$) and ${}^3P_2D$ ($c$). Right panel: Argand plots
for the dominant partial-wave amplitudes ${}^1D_2P$ ($d$),
${}^3F_3D$ ($e$) and ${}^3P_2D$ ($f$). Dash-dotted lines show the
summed contributions of two conventional mechanisms ONE +
$N\Delta$ with a cut-off parameter $\tilde{\Lambda}_* =0.3$ GeV
consistent with $\pi N$ elastic scattering (see Fig.~\ref{fig2}).
The contributions of ONE + $N\Delta$ mechanisms with an enhanced
parameter $\tilde{\Lambda}_* =0.52$ GeV are shown by dashed lines.
The ONE + $N\Delta$ contributions in the ${}^3P_2D$ channel were
multiplied by a factor of 5 for better visibility. Results of the
full model calculations including also intermediate dibaryon
resonances are shown by solid lines. The open circles and dotted
lines correspond to the PWA results (SAID, solutions C500 and
SP96, respectively~\cite{SAID,Oh97}).} \label{fig3}
\end{figure}

Nevertheless, in determining these parameters, one should bear in
mind the possible uncertainties associated with our model
assumptions as well as with different PWA results. Thus, two SAID
PWA solutions, i.e., the coupled-channel solution
C500~\cite{SAID,Oh97} and the previous solution SP96~\cite{SAID}
for the $\pi^+ d \rightarrow pp$ reaction, give almost similar
results for two dominating partial-wave amplitudes, but rather
different results for the smaller ${}^3P_2D$ amplitude at energies
$\sqrt{s}
> 2.17$ GeV --- see Fig.~\ref{fig3}. Our present fit for the latter
amplitude gives some average result between these PWA solutions.
Further, in view of a large width of the ${}^3P_2$ resonance, the
extracted values of its parameters depend on the width
parameterization used in the theoretical model. Thus, assuming the
total width to be constant, we obtained for this resonance $M_D =
2162$ and $\Gamma_D = 154$ MeV. These values almost coincide with
those found in~\cite{Arndt87} in the PWA of $pp$ elastic
scattering.

\begin{table*}[!ht]
\caption{\label{table1}Parameters of dibaryon resonances used in
calculations of the reaction $pp \to d \pi^+$. In the last column,
the phases $\varphi$ of the dibaryon production amplitudes with
respect to the ``background'' (ONE + $N\Delta$) amplitude are
shown.}
\begin{ruledtabular}
\begin{tabular}{ccccccc}
  ${}^{2S+1}L_J$ & $M_D$ (MeV) & $\Gamma_D$ (MeV) & $\Gamma_i \Gamma_f$ (MeV$^2$) &
  $\alpha_{pp}$ (GeV) & $\Lambda_{\pi d}$ (GeV) & $\varphi$ (deg)
\\ \hline ${}^1D_2$ & 2155 & 101 & 74 & 0.23
& 0.27 & 0 \\ ${}^3F_3$ & 2197 & 152 & 53 & 0.32 & 0.53 & 0 \\
${}^3P_2$ & 2211 & 195 & 450 & 3.0 & 0.26 & 180
\\
\end{tabular}
\end{ruledtabular}
\end{table*}

The results for the partial cross sections in three dominant
partial waves are shown in the left panel of Fig.~\ref{fig3}. The
Argand plots for the respective amplitudes\footnote{Note that due
to a different normalization, partial-wave amplitudes defined in
Sect.~\ref{sec2} should be multiplied by a factor
$\sqrt{pq/s}/8\pi$ to be compared with the ones used in the SAID
PWA~\cite{Oh97}.} are given in the right panel of Fig.~\ref{fig3}.

As is seen from Fig.~\ref{fig3}, the conventional mechanisms give
approximately $40$--$50$\% of the cross sections in the partial
waves ${}^1D_2P$ and ${}^3F_3D$. One should note that the initial-
and final-state distortions which are not included in our model,
would further decrease the calculated cross sections by about
20\%~\cite{Grein84}. On the other hand, the $N\Delta$ attraction
generated by pion exchange can enhance the cross sections
somehow~\cite{Niskanen95}. We argue that this $t$-channel
attraction which is governed by the cut-off parameters in the
meson-baryon vertices should be very moderate when using soft
cut-off values, and the basic short-range attraction in the
$N\Delta$ system would be induced in this case by generation of
intermediate dibaryon resonances, similarly to that found for the
$NN$ system in the dibaryon model for $NN$
interaction~\cite{JPG01K,IJMP02K}. Nevertheless, to examine
possible effects of the $N\Delta$ $t$-channel interaction on the
$pp \to d \pi^+$ cross sections, one could strengthen the
intermediate $\Delta$ contribution through enhancing the cut-off
parameter $\tilde{\Lambda}_*$ in the $\pi N \Delta$ vertex
\textit{ad hoc}. Thus, when enhancing $\tilde{\Lambda}_*$ from
$0.3$ to $0.52$ GeV, one is able to reproduce the magnitude of the
${}^1D_2P$ partial cross section (see Fig.~\ref{fig3}$(a)$). Note,
however, that this worsens simultaneously the description of
$P_{33}$ $\pi N$ elastic scattering (cf. Fig.~\ref{fig2}).
Further, as is shown in Fig.~\ref{fig3}$(b)$, the same cut-off
parameter modification can also improve the description of the
${}^3F_3D$ partial cross section, though not enough to reproduce
its empirical behavior.

On the other hand, in the ${}^3P_2D$ channel, the ONE + $N\Delta$
mechanisms give only ca. $2.5$\% of the partial cross section near
the resonance peak, and this result very weakly depends on the
$\pi N \Delta$ cut-off parameter value (see Fig.~\ref{fig3}$(c)$).
So that, the intermediate $\Delta$ excitation appears to play only
a minor role in this channel. Besides that, the $N\Delta$
amplitude has an improper phase here (see Fig.~\ref{fig3}$(f)$).
So, the satisfactory description of the empirical data on the
${}^3P_2D$ partial cross section cannot be attained through any
changes in parameters of the conventional mechanisms, and an
additional resonance contribution appears to be urgently needed!
The crucial role of proper description of the ${}^3P_2D$
partial-wave amplitude is particularly seen in the
spin-correlation parameters, which will be discussed in the next
section.

\begin{figure}[!ht]
\begin{center}
\resizebox{0.6\columnwidth}{!}{\includegraphics{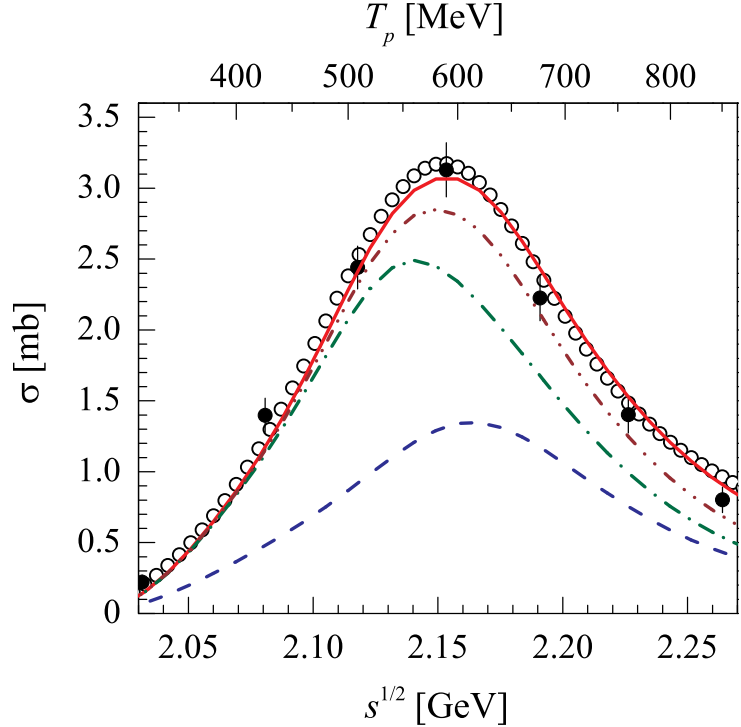}}
\end{center}
\caption{Total cross section of the reaction $pp \to d \pi^+$.
Dashed lines show the summed contributions of two conventional
mechanisms ONE + $N\Delta$. Dash-dotted, dash-dot-dotted and solid
lines correspond to the results of model calculations including
also one (${}^1D_2$), two (${}^1D_2 + {}^3F_3$) and three
(${}^1D_2 + {}^3F_3 + {}^3P_2$) intermediate dibaryon resonances,
respectively. Open circles correspond to PWA results (SAID,
solution C500~\cite{SAID,Oh97}) and filled circles
--- to the experimental data~\cite{Shimizu82}.} \label{fig4}
\end{figure}

Here, we present also the results for the total cross section
shown in Fig.~\ref{fig4}. Though the conventional mechanisms
reproduce a correct shape of the total cross section, the
experimental data are underestimated by a factor of two, similarly
to the partial cross sections in two dominant partial waves.
Taking the intermediate dibaryons into account fills in the
discrepancy between the conventional-model calculations and the
data, thus leading to very good reproduction of experimental data
in the whole energy range considered.

\section{Results for differential cross section and polarization
observables at $\bm{T_p=582}$~MeV} \label{sec4}

In this section, the results for differential observables in the
reaction $pp \to d \pi^+$ at energy $T_p=582$ MeV ($\sqrt{s}=2.15$
GeV) are presented. The energy value chosen here is close to that,
where the total cross section has its maximum. Besides that, a
rich set of experimental data exists in this energy
region~\cite{Hoftiezer83,Hoftiezer84,Smith84,Aprile82,Aprile84}.

\begin{figure}[!ht]
\begin{center}
\resizebox{0.6\columnwidth}{!}{\includegraphics{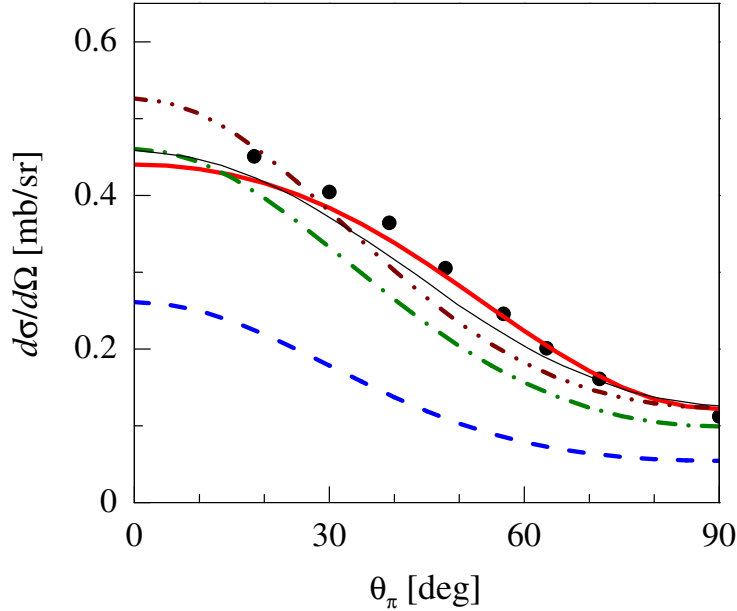}}
\end{center}
\caption{Differential cross section in the reaction $pp \to d
\pi^+$ at energy $T_p = 582$ MeV ($\sqrt{s}=2.15$ GeV). The
meaning of theoretical curves is the same as in Fig.~\ref{fig4}.
Thin solid line correspond to the conventional model
calculations~\cite{Lamot87} including off-shell modifications and
heavy-meson exchanges. Filled circles show the experimental
data~\cite{Hoftiezer83}.} \label{fig5}
\end{figure}

\begin{figure}[!ht]
\begin{center}
\resizebox{1.0\columnwidth}{!}{\includegraphics{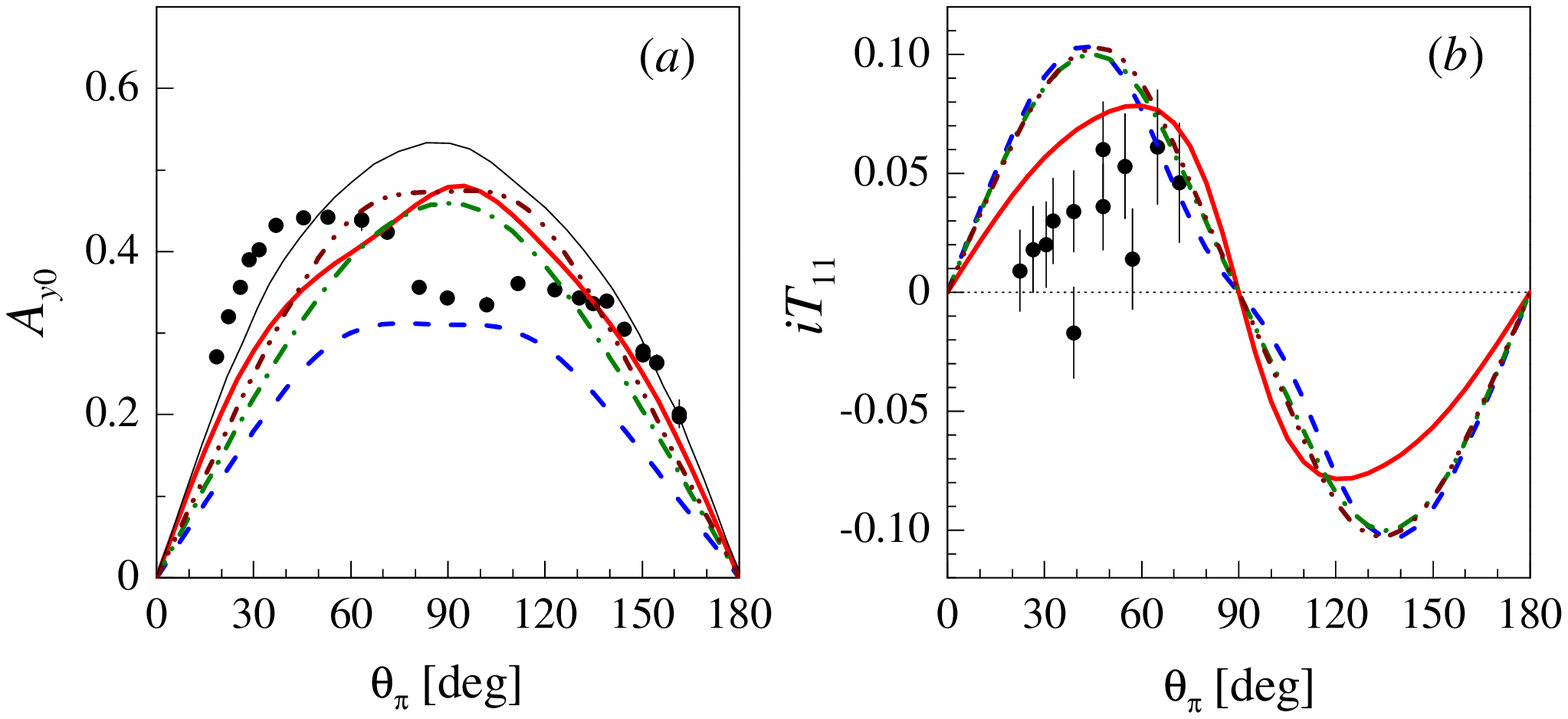}}
\end{center}
\caption{Proton ($a$) and deuteron ($b$) vector analyzing powers
in the reaction $pp \to d \pi^+$ at energy $T_p = 582$ MeV
($\sqrt{s}=2.15$ GeV). The meaning of theoretical curves is the
same as in Fig.~\ref{fig5}. Filled circles show the experimental
data~\cite{Hoftiezer84} ($a$) and~\cite{Smith84} ($b$). (The
$iT_{11}$ data~\cite{Smith84} were obtained in an inverse process
$\pi^+ d \to pp$ at $T_{\pi} = 140$~MeV, corresponding to $T_p =
562$~MeV.)} \label{fig6}
\end{figure}

The results for the differential cross section are shown in
Fig.~\ref{fig5}. The differential cross section is described very
well, except for the forward region, where high partial waves
(giving a contribution less than 3\% to the total cross section)
obviously play an important role. Thus, one can see some
underestimation of the contributions of these high partial waves
in our model. Further, the main contribution after the ${}^3P_2D$
channel comes from $S$-wave pion production in the ${}^3P_1S$
partial wave~\cite{Arndt93}. The $S$-wave pion production, which
is important near the threshold, is usually described by some
additional mechanism based on phenomenological Lagrangian
approach~\cite{Brack77}. Some contribution to this term comes also
from $S$-wave $\pi N$ scattering in the intermediate state. Both
these mechanisms are not included in the present model. This is
also the possible reason for a discrepancy between our calculation
and experimental data for the proton analyzing power $A_{y0}$
shown in Fig.~\ref{fig6}$(a)$. This observable is very sensitive
to the small amplitudes in the non-dominant partial waves, and
especially, to the ${}^3P_1S$ amplitude. In fact, just a few
models were able to reproduce the shape of $A_{y0}$ (see,
e.g.,~\cite{Grein84,Niskanen84}). It is so sensitive to the tiny
details of the model, that even the most accurate to date
theoretical calculation based on solving exact Faddeev-type
equations for the coupled $\pi NN \leftrightarrow NN$
system~\cite{Lamot87} could not reproduce its proper behavior. In
particular, as was shown in~\cite{Mizutani81}, inclusion of the
small $S$- and $P$-wave $\pi N$-scattering amplitudes just in
first order leads to a proper description of the ``double-hump''
shape of $A_{y0}$, but the exact inclusion of these small
amplitudes gives again an improper behavior like that shown in
Fig.~\ref{fig6}$(a)$. So, even the qualitative description of
$A_{y0}$ requires an extremely accurate theoretical treatment of
small partial-wave amplitudes.

We also calculated the deuteron vector analyzing power $iT_{11}$
(see Fig.~\ref{fig6}$(b)$). Its qualitative behavior is described
properly already by conventional mechanisms, however, with a
significant overestimation. Inclusion of dibaryon resonances,
especially the ${}^3P_2$ one, allows to reduce the discrepancy
with experimental data.

In general, one can conclude that our approximate model for
conventional meson-exchange processes describe the experimental
data for the basic observables not worse than more sophisticated
theoretical models elaborated in previous years. The main
difference is that we used soft values for the short-range cut-off
parameters in meson-baryon vertices (\textit{consistent with $\pi
N$ scattering}) and obtained lower cross sections, than other
models which used larger cut-off values fitted ad hoc to describe
the magnitude of the $pp \to d \pi^+$ cross section
(see~\cite{Lamot87}, Sect.~III --- ``The off-shell
modification''). Inclusion of dibaryon resonances is able to give
the lacking short-range contributions to the cross sections and
vector analyzing powers, however, without changing their
qualitative behavior.

\begin{figure}[!ht]
\begin{center}
\resizebox{1.0\columnwidth}{!}{\includegraphics{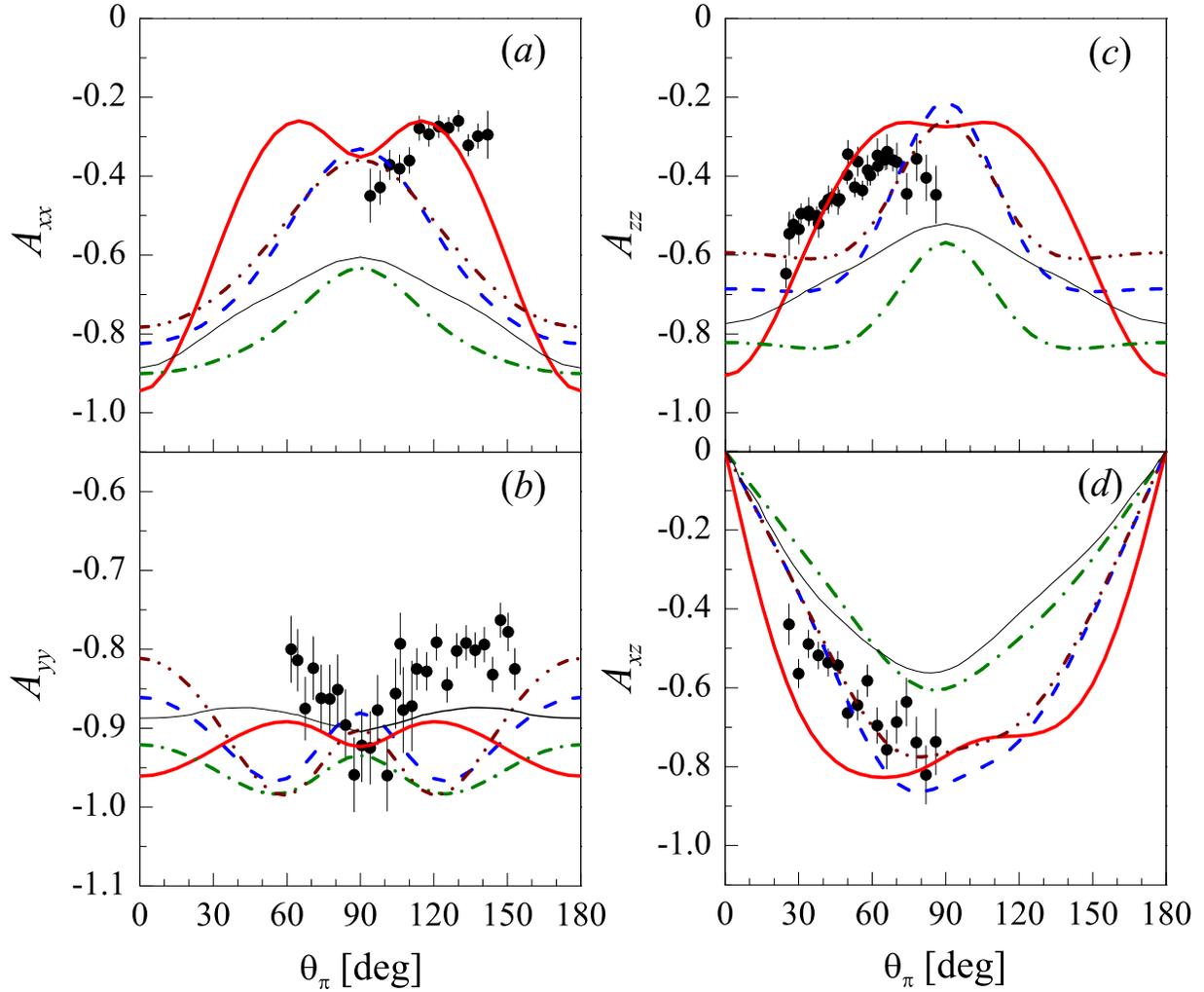}}
\end{center}
\caption{Spin-correlation parameters $A_{xx}$ ($a$), $A_{yy}$
($b$), $A_{zz}$ ($c$), and $A_{xz}$ ($d$) in the reaction $pp \to
d \pi^+$ at energy $T_p = 582$ MeV ($\sqrt{s}=2.15$ GeV). The
meaning of theoretical curves is the same as in Fig.~\ref{fig5}.
Filled circles show the experimental data~\cite{Aprile82,Aprile84}
for $T_p = 578$~MeV.} \label{fig7}
\end{figure}

The quite opposite situation takes place with the spin-correlation
parameters, the results for which are shown in Fig.~\ref{fig7}. It
is well-known that description of spin-correlation parameters in
the $pp \to d \pi^+$ reaction was a serious problem for
conventional theoretical models~\cite{Lamot87,Grein84,Niskanen84}.
It was established long ago~\cite{Grein84} that conventional
meson-exchange mechanisms underestimate the contributions of
triplet $pp$ partial waves, which are very important for
reproducing correctly the spin-correlation parameters. However, no
definite solution for this problem has been found previously. As
is seen from Fig.~\ref{fig7}, just the ${}^3P_2D$ amplitude and
its interference with other amplitudes changes strongly the
qualitative behavior of the spin-correlation parameters, thus
turning them into qualitative (or even semiquantitative) agreement
with experimental data. And the proper magnitude of this amplitude
can be obtained only by assuming a triplet $P$-wave dibaryon
resonance excitation in addition to the conventional $\Delta$
excitation (see Figs.~\ref{fig3} ($c$) and ($f$)). This is likely
\textit{one of the most important results} of the present study.

In view of this result, we can suggest the possible reason for
improper behavior of spin-correlation parameters in the model
calculations~\cite{Kamo79,Kamo80} which also included dibaryon
resonances. Although the ${}^3P_2$ resonance was included in these
calculations, its parameters were not fitted individually, but
together with parameters of the more intensive ${}^1D_2$ and
${}^3F_3$ resonances, to describe experimental data with mixed
contributions of all resonances. As a result, its mass was found
to be 2110 MeV and width 30 MeV, which are too low in comparison
with experimental values and our results.

\section{Discussion and summary}
\label{sec5}

In the paper, we studied the manifestation of isovector dibaryon
resonances ${}^1D_2$, ${}^3F_3$ and ${}^3P_2$ in the basic
single-pion production reaction $pp \to d \pi^+$. All these
resonances have been found in the PWA of $pp$ elastic
scattering~\cite{Arndt87,Higuchi91}, however, the $P$-wave
diproton resonance, being the least intensive, received less
attention in literature than the $D$- and $F$-wave resonances. A
new experimental evidence of the ${}^3P_2$ dibaryon has appeared
just very recently~\cite{Komarov16} in the reaction $pp \to (pp)_0
\pi^0$ where the more intensive ${}^1D_2$ and ${}^3F_3$ resonances
are forbidden by angular momentum and parity conservation.

We have found the large effects of the ${}^3P_2$ diproton in the
spin-correlation parameters of the $pp \to d\pi^+$ reaction. In
fact, the conventional models~\cite{Lamot87,Grein84,Niskanen84}
(based mainly on the $t$-channel meson-exchange mechanisms) for
this reaction generally resulted in underestimation and even
improper behavior of the proton-proton spin-correlation parameters
$A_{xx}$, $A_{yy}$ and $A_{zz}$. We should note here that the
coupled-channel approach of Niskanen~\cite{Niskanen84} appeared to
be more successful than other conventional models for $pp \to
d\pi^+$ reaction, while still giving essential underestimation of
$A_{xx}$ and $A_{yy}$. However this approach turned out to
completely fail for a similar reaction $pp \to (pp)_0 \pi^0$, as
was discussed in detail in a recent experimental
paper~\cite{Komarov16}. Thus, the most recent experimental data in
this area are in a strong disagreement with the model predictions
of Niskanen both in the forward cross section and in energy
dependence. On the other hand, just these experimental
data~\cite{Komarov16} revealed existence of the ${}^3P_2$
dibaryon. So, it is quite reasonable to suggest that the ${}^3P_2$
dibaryon which is clearly seen in the reaction $pp \to (pp)_0
\pi^0$, should also manifest itself in a similar reaction $pp \to
d\pi^+$, however, not in the unpolarized cross section where the
dominant contributions are given by ${}^1D_2$ and ${}^3F_3$
dibaryons, but in more sensitive observables like spin-correlation
parameters. In fact, we have shown that only assuming the
contribution of the $P$-wave diproton resonance makes it possible
to explain semi-quantitatively the experimental data for these
observables. By the way, this explanation is rather similar to the
explanation of the proton polarization in the reaction
$d(\gamma,\vec{p})n$ at $E_{\gamma}=400$--$600$~MeV found long ago
in works~\cite{Kamae77,Kamae77-2}. The strong disagreement for the
outgoing proton polarization between the predictions of
conventional models and experimental data~\cite{Kamae77} could
only be explained by incorporation of the ${}^3D_3$ and ${}^3F_3$
intermediate dibaryon contributions. In this case one has another
example of a process where the large spin-dependent observables
could be explained only by assuming the intermediate dibaryon
resonances.

Another interesting question worth to be discussed in connection
with the $P$-wave diproton is a well-known puzzling behavior of
the elastic $NN$-scattering phase shifts in the triplet $P$ waves.
In fact, while the triplet ${}^3P_0$ and ${}^3P_1$ $NN$ phase
shifts (and also the singlet one ${}^1P_1$) clearly demonstrate
the short-range repulsive core behavior (with the core radius
$r_{\rm c} \simeq 0.9$~fm), the ${}^3P_2$ phase shifts are rising
up to 600 MeV (lab.) and do not display any features of the
repulsive core. However, in the conventional treatment of $NN$
interaction, the short-range central-force repulsion should be a
universal feature for all ${}^3P_J$, $J=0,1,2$. The puzzle has
been resolved in the conventional OBE-like
models~\cite{BohrMottelsonI99} through introduction of a highly
intensive short-range spin-orbit force which produces a very
strong attraction just in the ${}^3P_2$ channel and compensates
completely the very large and broad repulsive core which is
present in all $P$ waves. This huge spin-orbit interaction looks
rather unnatural and fitted \textit{ad hoc} (for a detailed
discussion of inconsistencies in the OBE-like $NN$-potential
models see~\cite{PAN13}).

The results of the present paper give some alternative explanation
for the $P$-wave $NN$ phase-shifts puzzle. In fact, one can think
that the short-range ${}^3P_2$ dibaryon with a mass $M({}^3P_2)
\simeq 2.2$ GeV induces as usually a strong $NN$
attraction~\cite{JPG01K}, the strongest at lab. energies $T_{N}
\simeq 600$ MeV, so that, the above dibaryon-induced attraction at
intermediate energies can explain naturally the puzzling behavior
of the ${}^3P_2$ $NN$ phase shifts.

\begin{figure}[!ht]
\begin{center}
\resizebox{0.3\columnwidth}{!}{\includegraphics{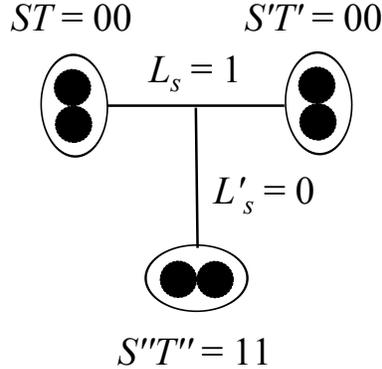}}
\end{center}
\caption{Allowable three-diquark configuration for an isovector
$P$-wave dibaryon.} \label{fig8}
\end{figure}

It is interesting to discuss further the possible quark structure
of an isovector $P$-wave dibaryon and a mechanism of its decay
with a pion emission. In the paper~\cite{NPA16} we adopted the
two-cluster $q^4 - q^2$ structure~\cite{Mulders80,Kondr87} for the
series of isovector dibaryons ${}^1D_2$, ${}^3F_3$, ${}^1G_4$,
etc., with a tetraquark $q^4(S=1,T=0)$ and an axial diquark
$q^2(S'=T'=1)$ connected by a color QCD string with an orbital
angular momentum $L_s=0,1,2$, etc. The ${}^3P_2$ isovector
dibaryon apparently does not belong to this series. If to assume
the two-cluster $q^4 - q^2$ structure for this dibaryon as well,
then the most appropriate structure would be a two-cluster state
with a color string ($L_s=1$) connecting the tetraquark
$q^4(S=T=1)$ and a scalar diquark $q^2(S'=T'=0)$. However the
tetraquark with $S=T=1$ should be unstable against the decay into
two (scalar and axial) diquarks. Then, two scalar diquarks
resulted from such decay into the three-diquark system must be in
a mutual $P$-wave due to the Pauli exclusion principle (see
Fig.~\ref{fig8}).

The ``natural'' decay of the three-diquark state with a pion
emission, viz. $q^2(S''=T''=1) \to q^2(S''=T''=0)$ and $L_s'=0 \to
L_s'=1$ is unlikely because the final state in this case would be
the so-called ``demon deuteron''~\cite{Fredriksson82} with two
$P$-wave strings, which would have a higher mass than the initial
$P$-wave dibaryon. Thus, the most probable transition should be
the rearrangement of the three-cluster configuration shown in
Fig.~\ref{fig8} to a conventional two-cluster $q^3 - q^3$ state
with a $P$-wave string between two $3q$ clusters. Then this state,
in its turn, decays via a single-pion emission to the final
deuteron $(S=1,T=0)$ or singlet deuteron $(S=0,T=1)$, i.e., by the
usual spin-flip or isospin-flip transitions.

The recent experiment~\cite{Komarov16} revealed existence of the
${}^3P_0$ dibaryon resonance, along with the ${}^3P_2$ one.
Besides that, the recent $pp$-scattering analysis~\cite{Vary15}
predicts three diproton resonances ${}^3P_J$, $J=0,1,2$. The decay
of the ${}^3P_0$ resonance into $d \pi^+$ channel is forbidden due
to angular momentum and parity conservation, but it can decay into
$(pp)_0 \pi^0$ channel. For the same reasons, the ${}^3P_1$
resonance can decay into $d \pi^+$, but not into $(pp)_0 \pi^0$
channel. If the ${}^3P_0$ and/or ${}^3P_1$ dibaryons really exist,
all the above quark-structure consideration will hold for them
either. However, these resonances obviously give a very small
contribution to $NN$ elastic scattering (compared to the
background meson-exchange mechanisms), which is indicated by a
very moderate attraction in the ${}^3P_0$ channel and an almost
negligible attraction in the ${}^3P_1$ channel (compared to the
strong attraction in the ${}^3P_2$ channel --- see the above
discussion of the $P$-wave phase-shifts puzzle). Furthermore,
these resonances, contrary to the ${}^3P_2$ dibaryon, were not
found in most phase-shift analyses of $pp$ elastic scattering. So,
further studies are needed to shed light on existence and
properties of the $P$-wave diproton resonances.

\textit{To summarize,} we have shown that intermediate dibaryon
resonances in the $NN$ channels ${}^1D_2$, ${}^3F_3$ and ${}^3P_2$
are very likely to be responsible for a significant part of the
cross sections of the basic single-pion production process $pp \to
d\pi^+$ in a broad energy range ($T_p=400$--$800$~MeV). Moreover,
the ${}^3P_2$ diproton resonance has been shown to be responsible
for the most important characteristic features of the $pp$
spin-correlation parameters in this reaction (at least near $T_p
\simeq 600$~MeV). So, the role of the isovector dibaryons in
single-pion production in $pp$ collisions is rather similar to
that of the isoscalar ${}^3D_3$ dibaryon in double-pion production
in $pn$ collisions~\cite{Adl11,PRC13}. Besides that, the isovector
dibaryons might play an important role also in double-pion
production in $pp$ collisions~\cite{NPA16}. These results may have
many far-going implications in hadronic and nuclear physics.

\section*{ACKNOWLEDGEMENTS} The work was done under partial financial support
from RFBR grants Nos.~16-02-00049, 16-02-000265 and 16-52-12005.
M.N.P. also appreciates support from Dynasty Foundation.

\bibliography{mybibfile}

\end{document}